\documentclass[%
 reprint,
superscriptaddress,
 amsmath,amssymb,
 aps,
 prl,
 twocolumn
]{revtex4}

\usepackage{graphicx}
\usepackage{dcolumn}
\usepackage{bm}
\usepackage[T1]{fontenc}
\usepackage{braket}
\usepackage[utf8x]{inputenc}
\usepackage{easyReview}
\usepackage{soul}
\usepackage{mathtools}
\usepackage[capitalise]{cleveref}

\begin{document}

\title{Non-Hermitian skin effect induced by Rashba-Dresselhaus spin-orbit coupling}

\preprint{APS/123-QED}

\author{Pavel~Kokhanchik}
\affiliation{Institut Pascal, Université Clermont Auvergne, CNRS, Clermont INP, F-63000 Clermont–Ferrand, France}

\author{Dmitry~Solnyshkov}
\affiliation{Institut Pascal, Université Clermont Auvergne, CNRS, Clermont INP, F-63000 Clermont–Ferrand, France}
\affiliation{Institut Universitaire de France (IUF), 75231 Paris, France}

\author{Guillaume~Malpuech}
\affiliation{Institut Pascal, Université Clermont Auvergne, CNRS, Clermont INP, F-63000 Clermont–Ferrand, France}

\date{\today}

\begin{abstract}
1D chains with non-reciprocal tunneling realizing the non-Hermitian skin effect (NHSE) have attracted considerable interest in the last years whereas their experimental realization in real space remains limited to a few examples. In this work, we propose a new generic way of implementing non-reciprocity based on a combination of the Rashba-Dresshlauss spin-orbit coupling, existing for electrons, cold atoms, and photons, and a lifetime imbalance between two spin components. We show that one can realize the Hatano-Nelson model, the non-Hermitian Su-Schrieffer-Heeger model, and even observe the NHSE in a 1D potential well without the need for a lattice. We further demonstrate the practical feasibility of this proposal by considering the specific example of a photonic liquid crystal microcavity. This platform allows one to switch on and off the NHSE by applying an external voltage to the microcavity.
\end{abstract}

\maketitle

\emph{Introduction.} The last decades have been marked by a growing interest in non-Hermitian physics~\cite{bergholtz2021exceptional}, which appears reasonable due to the ubiquitous non-equilibrium nature of physical systems. Many phenomena without Hermitian analogs have been found. From the mathematical perspective, these are exceptional points and their nodal phases, non-Hermitian symmetry operators and classes~\cite{kawabata2019symmetry}, anomalous bulk-boundary correspondence~\cite{lee2016anomalous} (BBC), and related non-Hermitian skin effect~\cite{yao2018edge} (NHSE). From the physical point of view, these mathematical concepts became a foundation for unidirectional invisibility \cite{feng2013experimental}, enhanced sensitivity \cite{Wiersig2014,park2020symmetry,Wiersig2022}, high-performance lasing \cite{peng2016chiral}, etc.

\begin{figure}[tbp]
\includegraphics[width=0.95\columnwidth]{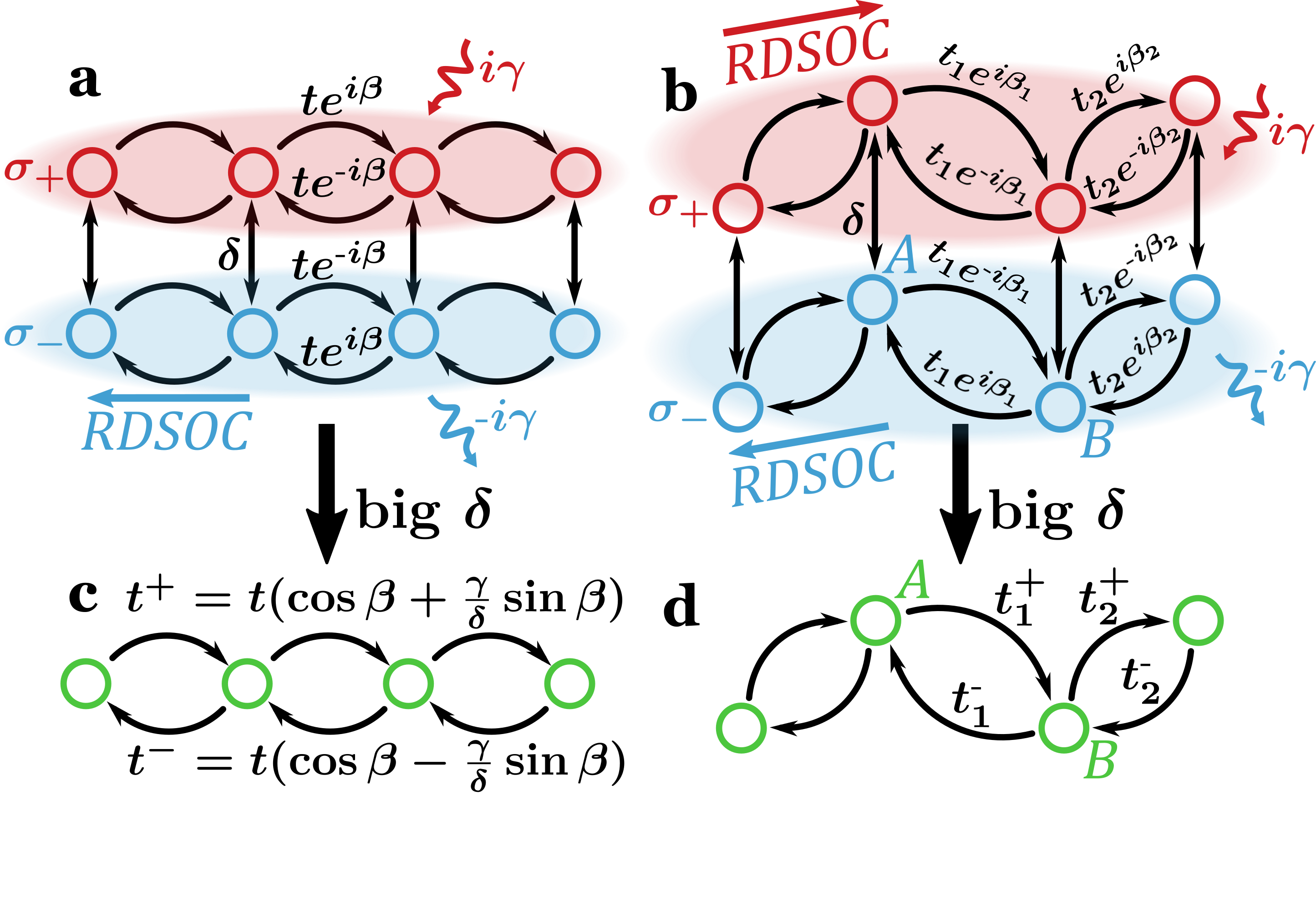}
\caption{(a) Spinful monomer and SSH chains with RDSOC, (effective) in-plane magnetic field and spin lifetime imbalance; (c) non-Hermitian HN and (d) SSH models.
\label{fig_0}}
\end{figure}

The BBC~\cite{Hatsugai1993,Hasan2010,ozawa2019topological} relates the topologically nontrivial structure of bulk states with the presence of topologically protected edge states. It relies on the assumption that the introduction of the boundary does not induce any effects on the bulk. While the BBC is valid in the Hermitian systems, it was found to be broken in many classes of non-Hermitian systems~\cite{lee2016anomalous,leykam2017edge,xiong2018does,martinez2018topological}. The BBC breakdown is always accompanied by the piling up of the bulk states of the system at the boundary (skin modes), the effect which was dubbed NHSE in analogy with the skin effect in electromagnetism~\cite{hatano1996localization,longhi2017nonadiabatic,kunst2018biorthogonal,yao2018edge}. The simplest model demonstrating NHSE is the Hatano-Nelson (HN) model~\cite{hatano1996localization}, a 1D chain with nonreciprocal couplings. However, the breakdown of BBC can not be observed in the HN chain since its Hermitian analog, a monomer chain, does not possess edge states. Thus the simplest model where the BBC breakdown was first investigated is the non-Hermitian Su-Schrieffer-Heeger (SSH) model (\cref{fig_0}b)~\cite{kunst2018biorthogonal,yao2018edge,lieu2018topological,yin2018geometrical}, a
non-Hermitian extension of 1D~SSH dimer chain possessing topologically protected edge states~\cite{su1979solitons,asboth2016short}.
The two main and complementary ways to restore the BBC were suggested: 
using special topological invariants for non-Bloch Hamiltonians of infinite systems~\cite{yao2018edge,yokomizo2019non} or calculating the biorthogonal polarization for finite systems with open boundary conditions (OBC) Hamiltonian~\cite{kunst2018biorthogonal,edvardsson2019non,edvardsson2020phase}.

The two frameworks exist as well to describe the NHSE: the first is based on the non-Hermitian winding number calculated for Bloch Hamiltonian with periodic boundary condition (PBC) (see \cref{NH_winding})~\cite{gong2018topological,shen2018topological}. The other utilizes the concept of the generalized Brillouin zone (GBZ)~\cite{yokomizo2019non,song2019non}. Even though the two approaches are equivalent~\cite{zhang2020correspondence}, combining both is helpful to visualize the complementary information about the system eigenstates and eigenenergies.

The set of theoretical papers establishing the fundamentals of NHSE has triggered several experimental implementations, mostly in acoustics~\cite{zhang2021observation_a,zhang2021acoustic}, mechanics~\cite{wang2022non,ghatak2020observation}, and electrical circuits~\cite{helbig2020generalized,hofmann2020reciprocal,liu2021non,zhang2021observation_b,zou2021observation}. Theoretical suggestions for NHSE systems in the field of photonics are based mostly on coupled ring resonators~\cite{longhi2015robust,zhu2020photonic,song2020two}, exciton-polariton lattices~\cite{mandal2020nonreciprocal,xu2021nonreciprocal,mandal2022topological,xu2022non}, and photonic crystals~\cite{zhong2021nontrivial,yokomizo2022non,fang2022geometry}. According to our knowledge, experimental realizations of NHSE in photonics involve only discrete-time quantum walks in coupled optical fiber loops~\cite{weidemann2020topological} and bulk optics~\cite{xiao2020non}, synthetic frequency dimension in optical ring resonator~\cite{wang2021generating,wang2021topological}, and single realization in a real space utilizing a chain of active ring resonators~\cite{liu2022complex}. Therefore, the field still lacks a generic theoretical proposal of NHSE in real space that could be applied to a large range of platforms.

In this work, we present a generic implementation of non-reciprocity and NHSE, not related to fabricating microstructures or complicated lattices, and achievable in several platforms. The approach is based on coupling between spin and propagation direction by Rashba-Dresselhaus spin-orbit coupling (RDSOC) (existing for electrons~\cite{bernevig2006exact,koralek2009emergence}, cold atoms~\cite{lin2011spin}, and photons~\cite{rechcinska2019engineering}) and in-plane effective magnetic field. By adding a spin-dependent lifetime imbalance (realized as well for electrons~\cite{weinelt2007spin}, cold atoms~\cite{tojo2009spin,ren2022chiral,qin2022imaginary} and photons~\cite{suppl}) we introduce non-Hermiticity into the system and obtain non-reciprocal couplings. This allows us to model the HN and non-Hermitian SSH chains. Furthermore, with these ingredients, we obtain the NHSE without any lattice, in a single 1D trap. In a particular case of photonics, RDSOC, in-plane magnetic field, and spin-dependent lifetime imbalance are equivalent to optical activity~\cite{ren2021nontrivial}, linear birefringence, and circular dichroism, respectively. The two first ingredients appear naturally in microcavities filled with liquid crystals (LCs)~\cite{rechcinska2019engineering}. The circular dichroism can be produced by embedding a chiral optical absorber~\cite{han2022broad,long2020chiral} or with spin-dependent gain~\cite{carlon2019optically}. We present realistic simulations showing NHSE using LC microcavity parameters. Finally, we show how the localization of skin modes can be controlled by an external voltage applied to LC microcavity, up to switching off the NHSE. 

\emph{1D monomer chain.} We start by considering the Hamiltonian of a spin-$\frac{1}{2}$ particle in a 1D monomer chain of coupled potential wells (sites), in the presence of RDSOC, in-plane (effective) magnetic field, and spin-dependent lifetime imbalance in the tight-binding approximation (\cref{fig_0}a):
\begin{equation}
\begin{split}
 \hat{H}_1 = \sum_{n} & \delta \hat{\sigma}_x a_{n}^{\dag} a_{n}^{} + i \gamma \hat{\sigma}_z a_{n}^{\dag} a_{n}^{} \\
 & + t e^{i \beta \hat{\sigma}_z} a_{n+1}^{\dag} a_{n}^{} + t e^{-i \beta \hat{\sigma}_z} a_{n}^{\dag} a_{n+1}^{},
 \end{split}
 \label{Ham_1}
\end{equation}
with $a_{n}^{} (a_{n}^{\dag})$ the annihilation (creation) operator of a particle at lattice site $n$, $\hat{\sigma}_i$ the $i$th Pauli matrix, $t$ the tunneling coefficient, $\delta$ an in-plane magnetic field. The RDSOC can be represented as a constant gauge potential that enters the tunneling coefficient as a spin-dependent phase $\beta \hat{\sigma}_z$ (see details in Ref.~\cite{kokhanchik2022modulated}), in analogy with the Harper-Hofstadter model~\cite{harper1955single,hofstadter1976energy} or Aharonov-Bohm effect~\cite{aharonov1959significance,aharonov1984topological}. The new ingredient here is the on-site spin lifetime imbalance which was absent in our previous work~\cite{kokhanchik2022modulated} and which makes the Hamiltonian non-Hermitian. 
The average lifetime is not included in the Hamiltonian and would only lead to a global decay of modes.

We first look at the limit of a large in-plane magnetic field $\delta$. The spin-subspace components of zero-order eigenstates are ones of the $\hat{\sigma}_x$ matrix (the first term in~\cref{Ham_1}), while their perturbation is defined by the remaining terms. In this case the effective Hamiltonian~\eqref{Ham_1} for $\ket{n} \otimes \ket{\sigma_{x};-}$ subspace reads:
\begin{equation}
\hat{H}_1^{eff} = \hat{H}_{1,onsite}^{eff} + \hat{H}_{1,NN}^{eff} + \hat{H}_{1,NNN}^{eff},
\label{Ham_1_eff}
\end{equation}
where
\begin{equation}
  \hat{H}_{1,NN}^{eff} = \sum_{n} t^- a_{n}^{\dag} a_{n+1}^{} + t^+ a_{n+1}^{\dag} a_{n}^{},
 \label{Ham_1_eff_NN}
\end{equation}
\begin{equation}
t^{\pm} = t (\cos \beta \pm \frac{\gamma}{\delta} \sin \beta).
\label{Ham_1_eff_tunnelings}
\end{equation}
$\hat{H}_{1,onsite}^{eff}, \hat{H}_{1,NN}^{eff}, \hat{H}_{1,NNN}^{eff}$ stand for onsite, nearest neighbor (NN) and next-nearest neighbor (NNN) Hamiltonians, respectively (see exact formulas in Ref.~\cite{suppl}). As one can see from~\cref{Ham_1_eff_NN,Ham_1_eff_tunnelings}, forward and backward tunneling coefficients $t^{\pm}$ differ by a factor $\frac{\gamma}{\delta} \sin \beta$. In this limit, the chain is equivalent to the non-reciprocal HN model up to the NNN tunnelings (\cref{fig_0}c).

\begin{figure}[tbp]
\includegraphics[width=0.95\columnwidth]{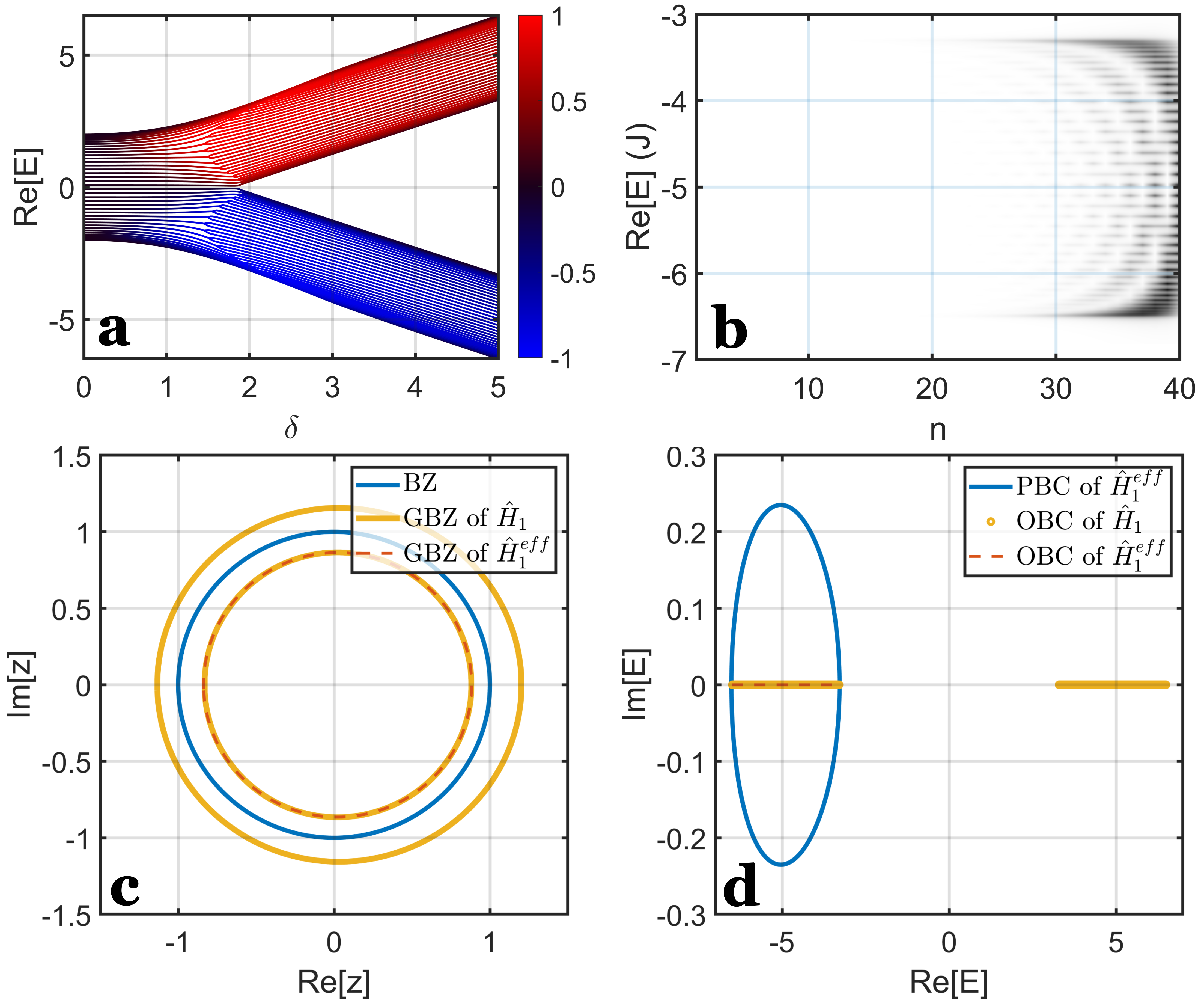}
\caption{(a) Transition from \cref{Ham_1} to \cref{Ham_1_eff} for a 40-site chain with an increase of $\delta$ displaying NHSE (blue (red) color shows the right (left) edge localization); (b) normalized density of eigenstates for the lower band of (a) depending on the real part of the energy $Re[E]$ and site number $n$; (c) BZ, GBZs of \cref{Ham_1,Ham_1_eff}; (d) PBC spectrum of \cref{Ham_1_eff}, OBC spectra of \cref{Ham_1,Ham_1_eff}; Parameters: $t=1, \gamma=1, \delta=5, \beta=0.2\pi$.
\label{fig_1}}
\end{figure}

We next consider a 40-site chain. \cref{fig_1}a shows the real eigenenergies of the full Hamiltonian~\eqref{Ham_1} versus $\delta$. With increasing $\delta$, a single band splits into two ($\ket{n} \otimes \ket{\sigma_{x};-}$ and $\ket{n} \otimes \ket{\sigma_{x};+}$ for big $\delta$). Each of the two bands shows a strong localization of right eigenstates (from here on just eigenstates) on the edge of the chain (opposite edges for different bands). \cref{fig_1}b demonstrates for $\delta=5$ the spatial distribution of modes of the lowest band, which confirms localization on the right edge. In a system where $\delta$ can be tuned experimentally, as in LC microcavities, NHSE can be turned on and off as shown by~\cref{fig_1}a.

As discussed in the introduction, the GBZ is a powerful tool to characterize the NHSE. In 1D Hermitian systems, the Brillouin zone (BZ) can be represented as the trigonometric circle $z=e^{ik}$, where $k$ is a real wavenumber. In non-Hermitian systems, it is appropriate to consider an imaginary wavenumber characterizing exponentially localized wavefunctions. This defines new eigenenergies where a part of the imaginary part is transferred to the wavenumber, resulting in the GBZ represented by $z=re^{ik}$. If a part of the GBZ plotted on a complex $z$ plane falls inside (outside) the BZ, corresponding eigenstates have $r<1 \ (r>1)$ and would therefore experience the NHSE with localization on the right (left) side of a finite chain. The GBZs of the full Hamiltonian~\eqref{Ham_1} and the effective Hamiltonian~\eqref{Ham_1_eff} are shown in \cref{fig_1}c by yellow and red lines, respectively. The GBZ of the full Hamiltonian~\eqref{Ham_1} consists of two closed lines corresponding to two bands localized on opposite edges. The GBZ of the reduced Hamiltonian~\eqref{Ham_1_eff} is in perfect agreement with the inner part of GBZ of full Hamiltonian~\eqref{Ham_1}, confirming their equivalence in this limit. In contrast to a pure HN model, red and yellow circles are not centered at $z=0$ due to the presence of NNNs (\cref{Ham_1_eff} and Ref.~\cite{suppl}).

The specificity of the non-Hermitian systems is that the spectra of infinite and finite systems differ profoundly. Indeed the use of periodic boundary conditions (PBC) requires considering only real wavenumbers, and the imaginary part is carried by the complex eigenenergies. \cref{fig_1}d depicts the OBC spectra of \cref{Ham_1} (yellow) and \cref{Ham_1_eff} (red) forming real value segments and complex PBC spectrum of \cref{Ham_1_eff} (blue). One can see that the effective Hamiltonian \eqref{Ham_1_eff} perfectly approximates the OBC spectrum of the lower energy band of full Hamiltonian \eqref{Ham_1}. The complex PBC spectrum can be used to compute non-Hermitian spectral winding number:
\begin{equation}
W_{E_b} \coloneqq \frac{1}{2\pi} \oint_{BZ} \frac{d}{dz} \arg [H(z) - E_b] dz, 
\label{NH_winding}
\end{equation}
with $H(z)$ PBC (Bloch) Hamiltonian, $E_b \in \mathbb{C}$ a reference point. NHSE, therefore, occurs if the loop of the PBC spectrum encircles a non-zero area, which is clearly the case for the blue line in \cref{fig_1}d.

\emph{SSH chain.} We continue by considering the SSH model -- the celebrated model in topological physics first suggested to describe the dimer structure of a polyacetylene chain~\cite{su1979solitons}. The chain contains two sublattices ($A$ and $B$), and the Hamiltonian is written as (\cref{fig_0}b):
\begin{equation}
\begin{split}
 & \hat{H}_2 = \sum_{n} \delta \hat{\sigma}_x (a_{n,A}^{\dag} a_{n,A}^{} + a_{n,B}^{\dag} a_{n,B}^{}) \\ 
 & + i \gamma \hat{\sigma}_z (a_{n,A}^{\dag} a_{n,A}^{} + a_{n,B}^{\dag} a_{n,B}^{}) + ( t_1 e^{i \beta_1 \hat{\sigma}_z} a_{n,B}^{\dag} a_{n,A}^{} \\
 & + t_2 e^{-i \beta_2 \hat{\sigma}_z} a_{n,B}^{\dag} a_{n+1,A}^{} + h.c. ),
 \end{split}
 \label{Ham_2}
\end{equation}
where $t_1 (t_2)$ intra-(inter-)cell tunneling with corresponding RDSOC phase $\beta_{1} (\beta_2)$, and all other parameters as in \cref{Ham_1}. By considering the $\ket{\sigma_x;-}$ subspace again, one can transform the Hamiltonian \cref{Ham_2} to:
\begin{equation}
\hat{H}_2^{eff} = \hat{H}_{2,onsite}^{eff} + \hat{H}_{2,NN}^{eff} + \hat{H}_{2,NNN}^{eff},
\label{Ham_2_eff}
\end{equation}
\begin{equation}
\begin{split}
  \hat{H}_{2,NN}^{eff} = \sum_{n} & t_1^+ a_{n,B}^{\dag} a_{n,A}^{} + t_1^- a_{n,A}^{\dag} a_{n,B}^{} \\
 & + t_2^- a_{n,B}^{\dag} a_{n+1,A}^{} + t_2^+ a_{n+1,A}^{\dag} a_{n,B}^{},
 \end{split}
 \label{Ham_2_eff_NN}
\end{equation}
\begin{equation}
t_i^{\pm} = t_i (\cos \beta_i \pm \frac{\gamma}{\delta} \sin \beta_i).
\label{Ham_2_eff_tunnelings}
\end{equation}

\begin{figure}[tbp]
\includegraphics[width=0.95\columnwidth]{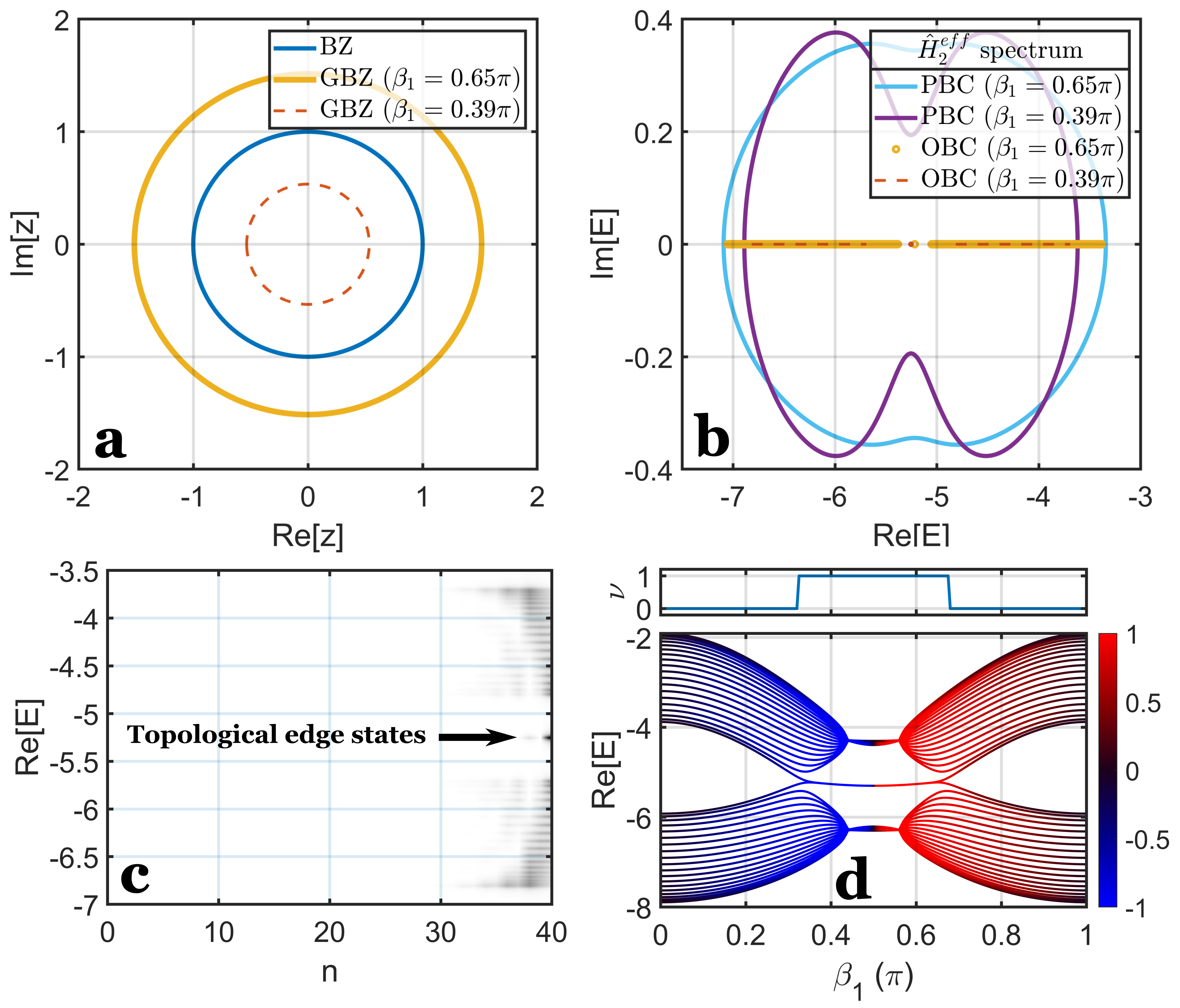}
\caption{(a) BZ, GBZs of \cref{Ham_2_eff} for two values of $\beta_1$; (b) PBC spectra of \cref{Ham_2_eff}, OBC spectra of \cref{Ham_2_eff} for two values of $\beta_1$, each showing two coinciding topological edge states; (c) normalized density of eigenstates of 40-site chain depending on the real part of the energy $Re[E]$ and site number $n$; (d) \cref{Ham_2_eff} real spectrum vs $\beta_1$ displaying NHSE (blue (red) color shows the right (left) edge localization) as well as a topological transition (bottom) confirmed by topological invariant $\nu$ (top). Parameters: $t_1=2, t_2=1, \gamma=1, \delta=5, \beta_2=0$.
\label{fig_2}}
\end{figure}

Next, we investigate the case when $\beta_2=0$ for simplicity (\cref{fig_0}d). Consequently, the NNN term vanishes (see Ref.~\cite{suppl}), and the chiral symmetry of Hamiltonian~\eqref{Ham_2_eff} is restored. We stress that the NHSE is present for other values of $\beta_2$ as well. The Hamiltonian \cref{Ham_2_eff} then can be transformed into the conventional SSH Hamiltonian  by similarity transformation $\hat{H}_{SSH}=\hat{S} \hat{H}_2^{eff} \hat{S}^{-1}$ with $\hat{S}=\text{diag} \{1,rs,r,r^2 s,...,r^{N-1},r^N s\}, r = \sqrt{t_1^- t_2^- / t_1^+ t_2^+}, s=\sqrt{t_1^+ / t_1^-}$, where $2N$ defines the number of lattice sites. The topology of the chain is then described by a Hermitian winding $\nu$ such that $\nu=1$ for $|t_2^+ t_2^-| > |t_1^+ t_1^-|$, and $\nu=0$ otherwise. The biorthogonal polarization invariant~\cite{kunst2018biorthogonal} shows the same transition for the initial non-Hermitian chain~\eqref{Ham_2_eff}. 

In \cref{fig_2}a we show GBZ for two different values of $\beta_1$: $\beta_1>\pi/2$ (yellow) and $\beta_1<\pi/2$ (red). They indicate the NHSE effect with accumulation on different edges, which is also confirmed by \cref{fig_2}d, demonstrating finite chain spectrum for different values of $\beta_1$ with a blue (red) color corresponding to the right (left) edge localization. The OBC spectra for the aforementioned values of $\beta_1$ are depicted in \cref{fig_2}b (yellow and red lines) together with PBC spectra (blue and purple lines). The individual dots between two bands of each OBC spectrum are the topological states. These topological states are localized at the edge of the chain (as the bulk states), but they are located inside the real gap of the OBC spectrum. They appear in \cref{fig_2}c, where we plot the normalized density of eigenstates versus the lattice site number. In the Hermitian limit ($\gamma \to 0$), the origin of the topological transition related to the formation of the edge states is the modulation of the tunneling amplitudes by the combination of RDSOC $\beta$ with the in-plane field $\delta$, as described in our previous paper~\cite{kokhanchik2022modulated}. Here, this transition is modified by non-Hermiticity and described by the invariant $\nu$ plotted in the top panel of \cref{fig_2}d.

\begin{figure}[tbp]
\includegraphics[width=0.95\columnwidth]{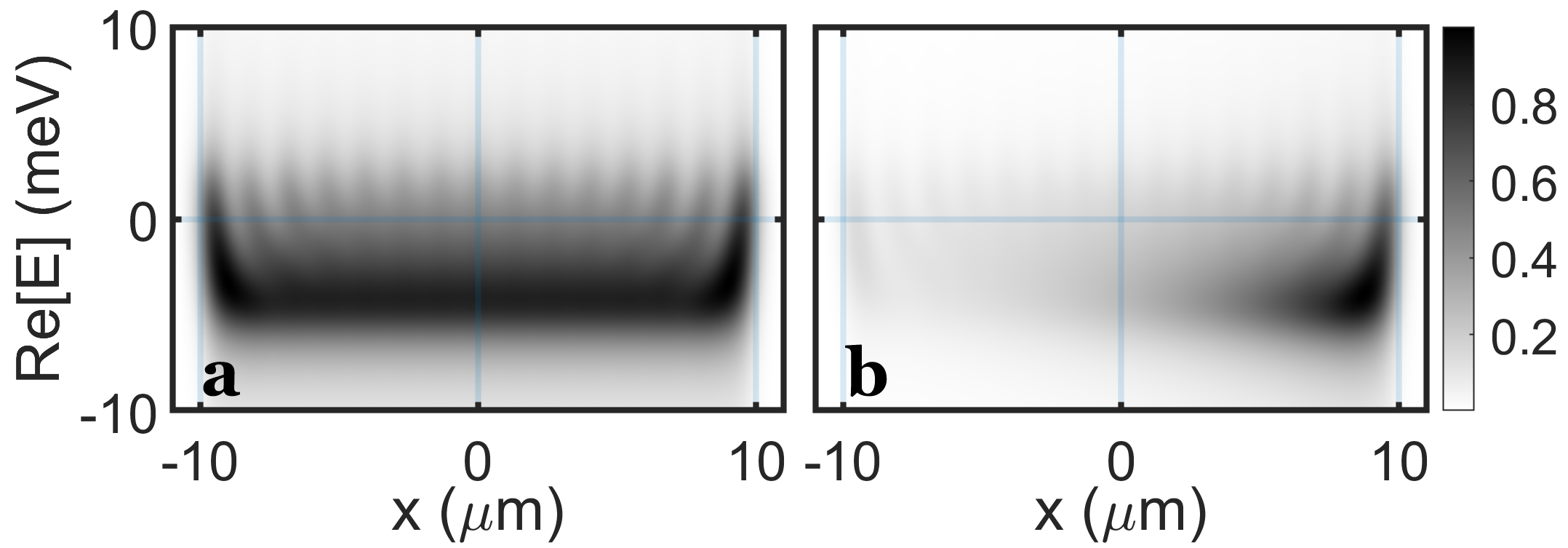}
\caption{Normalized eigenstate density for the lower energy states of the potential well depending on the real part of the energy $Re[E]$ and real space coordinate $x$. Parameters: $d=20 \ \mu \text{m}, \ U_0 = 10 \text{ meV}, \ m = 1.6 \cdot 10^{-5} m_e, \ m_e \text{ is an electron mass}, \ \alpha = 1 \text{ meV} \cdot \mu \text{m}, \ \delta=5 \text{ meV}, \ \gamma=0 \text{ meV}$ (a) and $\gamma=0.25 \text{ meV}$ (b).
\label{fig_3}}
\end{figure}

\emph{1D potential well.} Finally, we show that the NHSE 
can be observed in a potential well, without the use of a lattice. To do so, we solve numerically the 1D spinor stationary Schrödinger equation with the Hamiltonian:
\begin{equation}
\begin{split}
\hat{H}_{cont}(x) = & -\frac{\hbar^2}{2m} \frac{\partial^2}{\partial x^2} \hat{\sigma}_0 - 2i \alpha \frac{\partial}{\partial x} \hat{\sigma}_z \\ & + \delta \hat{\sigma}_x + i \gamma \hat{\sigma}_z + U(x) \hat{\sigma}_0,
\end{split}
\label{Ham_potwell}
\end{equation}
with $x$ the real space coordinate, $m$ the mass, $\gamma$ the spin lifetime imbalance, $\alpha$ the RDSOC magnitude. $U(x) = \{0, \text{ if } |x| < d; U_0, \text{ otherwise}\}$ is the real space potential with parameters $d = 20$~$\mu$m, $U_0 = 10$~meV. We plot the normalized eigenstate density for the Hermitian case $\gamma=0$~meV in \cref{fig_3}a and for the non-Hermitian case $\gamma=0.25$~meV in \cref{fig_3}b. The parameters used  correspond to realistic LC microcavities~\cite{rechcinska2019engineering}.
In particular, the broadening for every state is taken equal to $2$~meV~\cite{krol2022annihilation} in full-width half-maximum which exceeds the quantization energy, so the individual lines are indistinguishable. Nevertheless, it does not prevent a clear observation of NHSE: one can see that the non-Hermiticity drives the density of all lowest eigenstates to the right edge of the well. A similar effect was suggested for cold atoms~\cite{guo2022theoretical,li2022non}. The inversion of either the RDSOC direction ($\alpha \to -\alpha$) or of the spin lifetime imbalance ($\gamma \to -\gamma$) changes the NHSE so that the eigenstates localize at another edge. This property provides the tunability of the effect.

This effect can be explained again by considering one band $\ket{\sigma_x; -}$ in a limit of the large in-plane magnetic field $\delta$. Then, the Hamiltonian~\eqref{Ham_potwell} is rewritten as:
\begin{equation}
\hat{H}_{cont}^{eff}(x) = E_0 + U(x) - \xi^2 \left( \partial_x + \tau \right)^2,
\label{Ham_potwell_eff_gauge}
\end{equation}
with $E_0 = -\delta + \frac{\gamma^2}{2\delta} + (\frac{\alpha \gamma}{\xi \delta})^2$ constant energy shift, $\xi^2 = \frac{\hbar^2}{2m} - \frac{2\alpha^2}{\delta}$ kinetic energy scaling, $\tau = \frac{\alpha \gamma}{\xi^2 \delta}$ imaginary gauge potential~\cite{longhi2021non}. By considering wavefunction ansatz $\psi(x) = \varphi(x) e^{-\tau x}$, we arrive at a simple eigenvalue problem of a single spinless particle in a potential well with a wavefunction $\varphi(x)$:
\begin{equation}
\hat{H}_{cont}^{eff}(x) \varphi(x) = \left( E_0 + U(x) - \xi^2 \partial^2_x \right) \varphi(x).
\label{Ham_potwell_eff_no_gauge}
\end{equation}
As one can see, $\tau^{-1}$ is a localization length of a wavefunction $\psi(x)$ and its finite value is a manifestation of NHSE. It is achieved only when $\alpha,\gamma,\delta$ all have non-zero values, which confirms the necessity of each component in our model. The sign of $\tau$ is controlled by the combination of signs of $\alpha$ and $\gamma$.

As said above one of the possible experimental realizations of this proposal is to use a photonic microcavity filled with LCs~\cite{rechcinska2019engineering}. Here, the spin degree of freedom is represented by the photon polarization. The angle of the LC molecule director can be controlled via an external voltage. When the splitting between the modes of different linear polarization is big enough due to the rotation of LC molecules, two modes of opposite parity can come into resonance, giving rise to the RDSOC term in the Hamiltonian~\eqref{Ham_potwell}. A slight rotation of the LC director out of this resonance adds a splitting between these successive modes, entering the Hamiltonian as $\delta \hat{\sigma}_x$ (effective in-plane magnetic field). The presence of a chiral absorber or spin-dependent gain in a non-linear system can create a polarization lifetime imbalance. The potential well can be created by structuring the mirror of the microcavity. This system possesses the continuous tunability of $\delta$, as well as the possibility to switch $\alpha$ between zero and non-zero values. This allows one to tune the localization length of NHSE $\tau^{-1}$ or switch it on and off. 

\emph{Conclusion.} We propose a generic and original way to realize non-reciprocal tunnelings. It allows the implementation of models based on non-reciprocal couplings (HN model, non-Hermitian SSH, etc). The method is based on combining RDSOC, (effective) in-plane magnetic field, and spin lifetime imbalance. Furthermore, we demonstrate that the NHSE can be observed even without lattice. This proposal is feasible for many different platforms in solid-state physics, cold atoms, or photonics. In particular, we simulate realistic LC microcavity where NHSE can be tuned by voltage.

\begin{acknowledgments}
This work was supported by the European Union Horizon 2020 program, through a Future and Emerging Technologies (FET) Open research and innovation action under Grant Agreement No.~964770 (TopoLight), by the ANR project Labex GaNEXT (ANR-11-LABX-0014), and by the ANR program "Investissements d'Avenir" through the IDEX-ISITE initiative 16-IDEX-0001 (CAP 20-25). 
\end{acknowledgments}

\clearpage

\renewcommand{\thefigure}{S\arabic{figure}}
\setcounter{figure}{0}
\renewcommand{\theequation}{S\arabic{equation}}
\setcounter{equation}{0}

\section{Supplementary Materials}

\subsection{Effective Hamiltonian of a monomer chain in big $\delta$ limit}

\begin{equation}
\hat{H}_1^{eff} = \hat{H}_{1,onsite}^{eff} + \hat{H}_{1,NN}^{eff} + \hat{H}_{1,NNN}^{eff},
\label{Ham_1_eff}
\end{equation}
where
\begin{equation}
\hat{H}_{1,onsite}^{eff} = \sum_{n} \left( -\delta + \frac{\gamma^2 - 2 t^2 \sin^2 \beta}{2 \delta} \right) a_{n}^{\dag} a_{n}^{},
\label{Ham_1_eff_onsite}
\end{equation}
\begin{equation}
  \hat{H}_{1,NN}^{eff} = \sum_{n} t^- a_{n}^{\dag} a_{n+1}^{} + t^+ a_{n+1}^{\dag} a_{n}^{},
 \label{Ham_1_eff_NN}
\end{equation}
\begin{equation}
\hat{H}_{1,NNN}^{eff} = \sum_{n} \frac{t^2 \sin^2 \beta}{2 \delta} (a_{n}^{\dag} a_{n+2}^{} + a_{n+2}^{\dag} a_{n}^{}).
\label{Ham_1_eff_NNN}
\end{equation}
\begin{equation}
t^{\pm} = t (\cos \beta \pm \frac{\gamma}{\delta} \sin \beta).
\label{Ham_1_eff_tunnelings}
\end{equation}

\subsection{Effective Hamiltonian of a SSH chain in big $\delta$ limit}

\begin{equation}
\hat{H}_2^{eff} = \hat{H}_{2,onsite}^{eff} + \hat{H}_{2,NN}^{eff} + \hat{H}_{2,NNN}^{eff},
\label{Ham_2_eff}
\end{equation}
\begin{equation}
\begin{split}
\hat{H}_{2,onsite}^{eff} = & \sum_{n} \left( -\delta + \frac{\gamma^2 - (t_1^2 \sin^2 \beta_1 + t_2^2 \sin^2 \beta_2)}{2 \delta} \right) \times \\
& ( a_{n,A}^{\dag} a_{n,A}^{} + a_{n,B}^{\dag} a_{n,B}^{} ),
\end{split}
\label{Ham_2_eff_onsite}
\end{equation}
\begin{equation}
\begin{split}
  \hat{H}_{2,NN}^{eff} = \sum_{n} & t_1^+ a_{n,B}^{\dag} a_{n,A}^{} + t_1^- a_{n,A}^{\dag} a_{n,B}^{} \\
 & + t_2^- a_{n,B}^{\dag} a_{n+1,A}^{} + t_2^+ a_{n+1,A}^{\dag} a_{n,B}^{},
 \end{split}
 \label{Ham_2_eff_NN}
\end{equation}
\begin{equation}
\begin{split}
& \hat{H}_{2,NNN}^{eff} = \sum_{n} \frac{t_1 t_2 \sin \beta_1 \sin \beta_2}{2 \delta} ( a_{n+1,A}^{\dag} a_{n,A}^{} \\
& + a_{n+1,B}^{\dag} a_{n,B}^{} + a_{n,A}^{\dag} a_{n+1,A}^{} + a_{n,B}^{\dag} a_{n+1,B}^{} ),
\end{split}
\label{Ham_2_eff_NNN}
\end{equation}
\begin{equation}
t_i^{\pm} = t_i (\cos \beta_i \pm \frac{\gamma}{\delta} \sin \beta_i).
\label{Ham_2_eff_tunnelings}
\end{equation}

\subsection{Spin-dependent lifetime imbalance (circular dichroism)}

\begin{figure}[tbp]
    \centering
    \includegraphics[width=\linewidth]{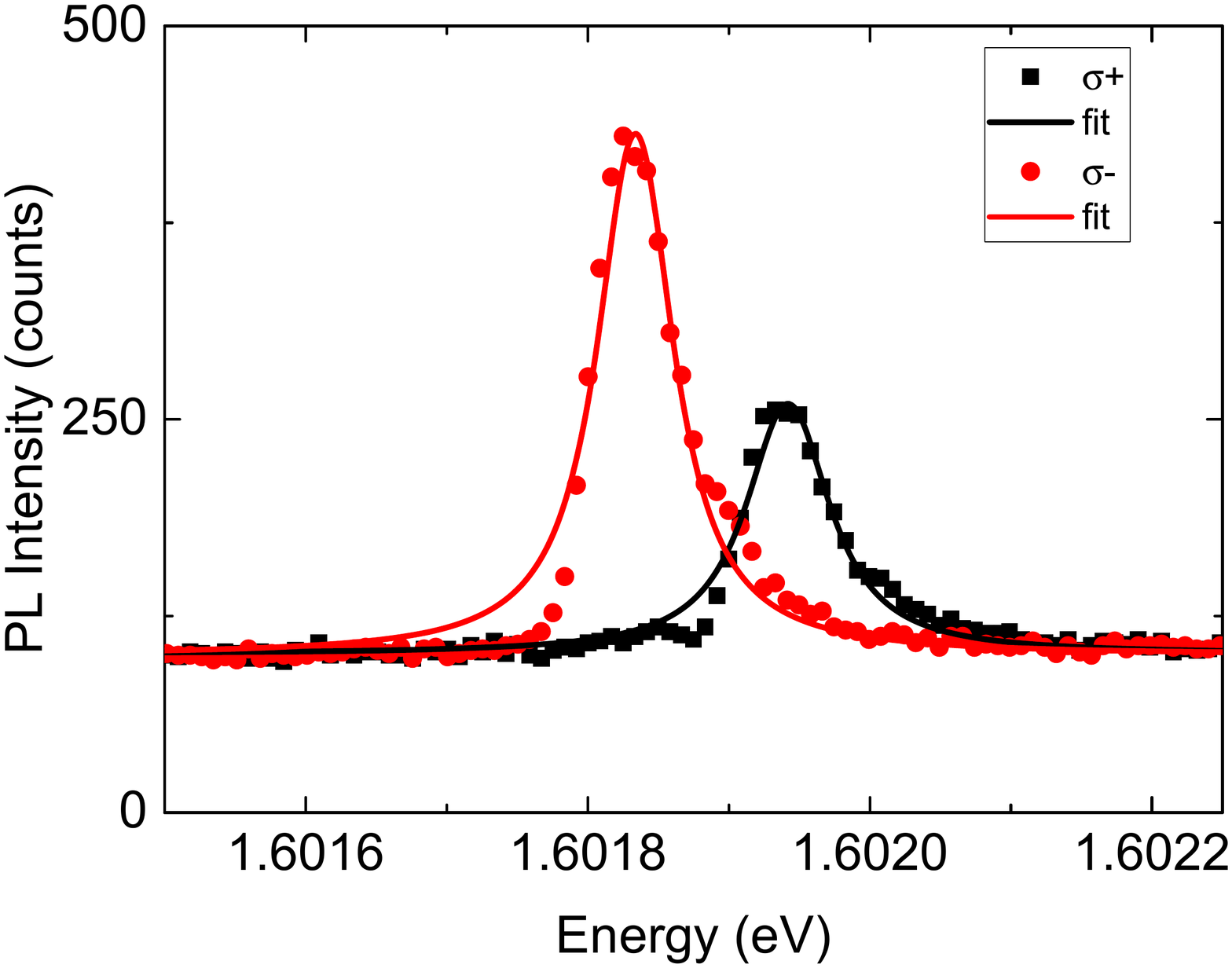}
    \caption{PL intensity from a GaAs microcavity~\cite{gianfrate2020measurement} at $k=0$ under applied magnetic field $B=9$~T in two circular polarizations with Lorentz fits demonstrating circular dichroism.}
    \label{FigS1}
\end{figure}

One of the key ingredients of the model presented in the main text leading to the non-Hermitian skin effect is the spin-dependent lifetime: the difference between the decay times, or equivalently the linewidths, of the two spins (polarizations). In optics, such an effect is often called circular dichroism. It may have different origins ranging from chiral absorption (due to the properties of the active material) to chiral pumping. Here, we present the analysis of the experimental data, publicly available in the Open Science Framework (OSF) repository associated with Ref.~\cite{gianfrate2020measurement} (\url{https://osf.io/s4rzu/?view_only=1cabd49416c04a9baed856dee3ae1ba9}). In this work, the authors have experimentally measured and analyzed the photoluminescence (PL) of a strongly-coupled GaAs microcavity under an applied magnetic field of $9$~T. The photoluminescence from polariton states occurs because these states are populated by scattering from the exciton reservoir. Under a strong magnetic field, the populations of two polarizations in the exciton reservoir are not equal, and the scattering towards the two circular-polarized polariton branches is therefore also different. The scattering into the state partially compensates its decay, leading to the reduction of its broadening. We extract the linewidths by fitting the measured PL intensity as a function of energy with a Lorentzian distribution function. The quality of the obtained fit was $R^2>0.94$. The asymmetry (non-zero third moment) of the experimental curves is due to the diffraction-induced contribution of the states at higher $k$, which all have higher energy than the state at $k=0$. We obtain a linewidth of $61\pm0.7$~$\mu$eV for $\sigma^+$ and $68\pm1$~$\mu$eV for $\sigma^-$, which corresponds to a difference of 11\%. The higher-energy state exhibits a shorter lifetime because of the smaller population of the associated exciton reservoir, as expected.

\bibliography{references}

\end{document}